
\input harvmac
\def\underarrow#1{\vbox{\ialign{##\crcr$\hfil\displaystyle
{#1}\hfil$\crcr\noalign{\kern1pt
\nointerlineskip}$\longrightarrow$\crcr}}}
\newcount\figno
\figno=0
\def\fig#1#2#3{
\par\begingroup\parindent=0pt\leftskip=1cm\rightskip=1cm\parindent=0pt
\baselineskip=11pt
\global\advance\figno by 1
\midinsert
\epsfxsize=#3
\centerline{\epsfbox{#2}}
\vskip 12pt
{\bf Fig. \the\figno:} #1\par
\endinsert\endgroup\par
}
\def\figlabel#1{\xdef#1{\the\figno}}
\def\encadremath#1{\vbox{\hrule\hbox{\vrule\kern8pt\vbox{\kern8pt
\hbox{$\displaystyle #1$}\kern8pt}
\kern8pt\vrule}\hrule}}

\overfullrule=0pt

%
\def\tilde{\widetilde}
\def\bar{\overline}

\font\zfont = cmss10 

\def\bigone{\hbox{1\kern -.23em {\rm l}}}
\def\ZZ{\hbox{\zfont Z\kern-.4emZ}}

\Title{hep-th/9411102, IASSNS-HEP-94-96}
{\vbox{\centerline{MONOPOLES AND FOUR-MANIFOLDS}}}
\smallskip
\centerline{Edward Witten}
\smallskip
\centerline{\it School of Natural Sciences, Institute for Advanced Study}
\centerline{\it Olden Lane, Princeton, NJ 08540, USA}\bigskip
\baselineskip 18pt

\medskip

\noindent
Recent developments in the understanding of $N=2$ supersymmetric Yang-Mills
theory in four dimensions suggest a new point of view about Donaldson
theory of four manifolds: instead of defining four-manifold invariants
by counting $SU(2)$ instantons, one can define equivalent four-manifold
invariants by counting solutions of a non-linear equation with an
abelian gauge group.  This is a ``dual'' equation in which the gauge
group is the dual of the maximal torus of $SU(2)$.
The new viewpoint suggests many new results about
the Donaldson invariants.
\Date{November, 1994}

\newsec{Introduction}
\nref\witten{E. Witten, ``Topological Quantum Field Theory,'' Commun. Math.
Phys. {\bf 117} (1988) 353.}
For some years now it has been known that Donaldson theory is equivalent
to a quantum field theory, in fact, a twisted version of $N=2$ supersymmetric
Yang-Mills theory \witten.
The question therefore arises of whether this viewpoint
is actually useful for computing Donaldson invariants \ref\doninv{S. Donaldson,
``Polynomial Invariants For Smooth Four-Manifolds,'' Topology, {\bf 29}
(1990) 257.} or understanding
their properties.

\nref\floer{A. Floer, ``An Instanton Invariant For 3-Manifolds,''
Commun. Math. Phys. {\bf 118} 215.}
One standard physical technique is to cut and sum over
physical states.  In the context of Donaldson theory, such methods
have been extensively developed by mathematicians,
starting with the work of Floer \floer.
So far, despite substantial efforts,
the physical reformulation has not given any essentially new insight
about these methods.

Another approach to using physics to illuminate Donaldson theory
starts with the fact that the $N=2$ gauge theory is
asymptotically free; therefore, it is weakly coupled in the ultraviolet
and strongly coupled in the infrared.  Since the Donaldson invariants
-- that is, the correlation functions of the twisted theory -- are metric
independent, they can be computed in the ultraviolet or the infrared,
as one wishes.  Indeed, the weak coupling in the
ultraviolet is used to show that the quantum field theory correlation
functions do coincide with the Donaldson invariants.

\nref\newwitten{E. Witten, ``Supersymmetric Yang-Mills Theory On A
Four-Manifold,'' J. Math. Phys. {\bf 35} (1994) 5101.}
If one could
understand the infrared behavior of the $N=2$ theory, one might get
a quite different description and, perhaps, a quite different way to
compute the Donaldson invariants.  Until recently, this line of thought was
rather hypothetical for general four-manifolds
since the infrared behavior of $N=2$ super Yang-Mills
theory in the strong coupling region was unknown.
Previous work along these lines was therefore limited to K\"ahler manifolds,
where one can reduce the discussion to the $N=1$ theory, whose infrared
behavior was known.  This led to an almost complete determination \newwitten\
of the Donaldson invariants of K\"ahler manifolds with $H^{2,0}\not= 0$.

\nref\sw{N. Seiberg and E. Witten,  ``Electric-Magnetic Duality,
Monopole Condensation, And Confinement In $N=2$ Supersymmetric
Yang-Mills Theory,'' Nucl. Phys. {\bf B426} (1994) 19,
``Monopoles, Duality, And Chiral Symmetry
Breaking In $N=2$ Supersymmetric QCD,'' hep-th/9408099,
to appear in Nucl. Phys. B.}
\nref\seiberg{N. Seiberg, ``The Power Of Holomorphy -- Exact Results
In $4d$ SUSY Field Theories,'' hep-th/9408013.}
The purpose of the present paper is to exploit recent work by Seiberg
and the author
\sw\ in which the infrared behavior of the $N=2$ theory was
determined using methods somewhat akin to methods that have shed light
on various $N=1$ theories (for a survey see \seiberg).
The answer turned out to be quite surprising: the
infrared limit of the $N=2$ theory in the ``strongly coupled'' region
of field space is equivalent to a  weakly coupled theory of abelian
gauge fields coupled to ``monopoles.''  The monopole theory is
dual to the original theory in the sense that, for instance, the
gauge group is the dual of the maximal torus of the original gauge group.

\nref\km{P. Kronheimer and T. Mrowka, ``Recurrence Relations And
Asymptotics For Four-Manifold Invariants,'' Bull. Am. Math. Soc. {\bf 30}
(1994) 215, ``Embedded Surfaces And The Structure Of Donaldson's
Polynomial Invariants,'' preprint (1994).}
\nref\arg{P. Argyres and A. Faraggi,  ``The Vacuum Structure And Spectrum
Of $N=2$ Supersymmetric $SU(N)$ Gauge Theory,'' hep-th/9411057.}
\nref\yank{A. Klemm, W. Lerche, S. Yankielowicz, and S. Theisen,
``Simple Singularities and $N=2$ Supersymmetric Yang-Mills Theory,''
hep-th/9411048.}
Since the dual theory is weakly coupled in the infrared,
everything is computable in that region, and
in particular for gauge group $SU(2)$,
one does get an alternative formulation of the usual Donaldson
invariants.  Instead of computing the Donaldson invariants by counting
$SU(2)$ instanton solutions, one can obtain the same invariants
by counting the solutions of the dual equations, which involve
$U(1)$ gauge fields and monopoles.\foot{In this paper, we only consider
Donaldson theory with
gauge group $SU(2)$ or $SO(3)$, but an  analogous dual description
by abelian gauge fields and monopoles will hold for
any compact Lie group, the gauge group of the dual theory being always
the dual of the maximal torus of the original gauge group.  For example,
most of the results needed to write the precise monopole equations for
$SU(N)$ have been obtained recently \refs{\arg,\yank}.}

This formulation makes manifest various properties of the Donaldson
invariants.  For instance, one can get new proofs of some of the
classic results of Donaldson theory; one gets
a new description of the basic classes of Kronheimer and Mrowka \km, and some
new results about them; one gets a new understanding
of the ``simple type'' condition for four-manifolds;
one finds new types of vanishing theorems
that severely limit the behavior of Donaldson theory on manifolds
that admit a metric of positive scalar curvature;
and  one gets a complete determination of the Donaldson invariants
of K\"ahler manifolds with $H^{2,0}\not= 0$, eliminating the
assumptions made in \newwitten\ about the canonical divisor.

It should be possible to justify
directly the claims sketched in this paper about the consequences
of the monopole equations even if the relation to Donaldson theory
is difficult to prove.  The reformulation may make the problems
look quite different
as the gauge group is abelian and the most relevant moduli spaces
are zero dimensional.
{}From a physical point of view the dual description via monopoles and abelian
gauge fields should be simpler than the microscopic $SU(2)$ description
since in the renormalization group sense
it arises by ``integrating out the irrelevant degrees of freedom.''

\nref\vw{C. Vafa and E. Witten, ``A Strong Coupling Test Of $S$-Duality,''
hep-th/9408074, to appear in Nucl. Phys. B.}
\nref\om{C. Montonen and D. Olive, Phys. Lett. {\bf B72} (1977) 117;
P. Goddard, J. Nuyts, and D. Olive, Nucl. Phys. {\bf B125} (1977) 1.}
\nref\sen{A. Sen, ``Strong-Weak Coupling Duality In Four Dimensional
String Theory,'' hep-th/9402002.}
The monopole equations are close cousins of equations studied in section two
of \vw; the reason for the analogy is that in each case one is studying
$N=2$ theories of hypermultiplets coupled to vector multiplets.
The investigation in \vw\ dealt with microscopic Montonen-Olive duality
\refs{\om,\sen}, while the duality in Donaldson theory \sw\ is a sort of
phenomenological analog of this.

The monopole equations, definition of four-manifold invariants,
and relation to Donaldson theory are stated
in section two of this paper.  Vanishing theorems are used in section three
to deduce some basic properties.  Invariants
of K\"ahler
manifolds are computed in section four.
A very brief sketch of the origin in physics
is in section five.  A fuller account of the contents of section five
will appear elsewhere \ref\nsw{N. Seiberg and E. Witten, to appear.}.

\newsec{The Monopole Equations}

Let $X$ be an oriented, closed four-manifold on which we pick a Riemannian
structure with metric tensor $g$.
$\Lambda^pT^*X$,  or simply $\Lambda^p$,
will denote the bundle of real-valued $p$-forms,
and $\Lambda^{2,\pm}$ will be  the sub-bundle of $\Lambda^2$ consisting
of  self-dual or anti-self-dual forms.

The monopole equations relevant to $SU(2)$ or $SO(3)$
Donaldson theory can be described
as follows.  If $w_2(X)=0$,
then $X$ is a spin manifold and one can pick
positive and negative spin bundles
$S^+$ and $S^-$, of rank two.  (If there is more than one spin structure,
the choice of a spin structure will not matter as we will ultimately
sum over twistings by line bundles.)  In that case, introduce a complex line
bundle $L$; the data in the monopole equation will be a connection $A$
on $L$ and a section $M$ of $S^+\otimes L$.  The curvature two-form
of $A$ will
be called $F$ or $F(A)$; its self-dual and anti-self-dual
projections will be called $F^+$ and $F^-$.

If $X$ is not spin, the $S^{\pm}$ do not exist,
but their projectivizations ${\bf P}S^{\pm}$ do exist (as bundles with fiber
isomorphic to ${\bf CP}^1$).  A ${\rm Spin}_c$ structure (which exists
on any oriented four-manifold \ref\hirz{F. Hirzebruch and H. Hopf,
``Felder von Flachenelementen in 4-dimensionalen Mannigfaltigkeiten,''
Math. Annalen {\bf 136} (1958) 156.})
can be described as a choice of
a rank two complex vector bundle, which we write as $S^+\otimes L$,
whose projectivization is isomorphic to ${\bf P}S^+$.  In this situation, $L$
does not exist as a line bundle, but $L^2$ does\foot{One might
be tempted to call this bundle $L$ and write the ${\rm Spin}_c$
bundle as $S^+\otimes L^{1/2}$; that amounts to assigning magnetic
charge $1/2$ to the monopole and seems unnatural physically.};
the motivation for
writing the ${\rm Spin}_c$ bundle as $S^+\otimes L$ is that the tensor
powers of this bundle obey isomorphisms suggested by the notation.
For instance, $(S^+\otimes L)^{\otimes 2}\cong L^2\otimes(\Lambda^0\oplus
\Lambda^{2,+})$.
The data of the monopole equation
are now a section $M$ of $S^+\otimes L$ and a connection on $S^+\otimes L$
that projects to the Riemannian connection on ${\bf P}S^+$.  The symbol
$F(A)$ will now denote $1/2$ the trace of the curvature form of $S^+\otimes L$.

Since $L^2$ is an ordinary line bundle, one has an integral
cohomology class
$x=-c_1(L^2)\in H^2(X,{\bf Z})$.  (The minus sign makes some
later formulas come out in a standard form.) Note that $x$ reduces modulo two
to $w_2(X)$; in particular, if $w_2(X)=0$, then $L$ exists as a line
bundle and $x=-2c_1(L)$.

To write the monopole equations, recall that $S^+$ is symplectic  or
pseudo-real, so that
if $M$ is a section of $S^+\otimes L$, then the complex conjugate $\bar M$
is a section of $S^+\otimes L^{-1}$.  The product $M\otimes \bar M$
would naturally lie in $(S^+\otimes L)\otimes (S^+\otimes L^{-1})\cong
\Lambda^0\oplus\Lambda^{2,+}$.
$F^{+}$ also takes values in $\Lambda^{2,+}$ making it possible to
write the following equations.
Introduce Clifford matrices $\Gamma_i$
(with anticommutators $\{\Gamma_i,\Gamma_j\}=2g_{ij}$), and
set $\Gamma_{ij}={1\over 2}[\Gamma_i,\Gamma_j]$. Then
the equations are\foot{
To physicists the connection form $A$ on a unitary line bundle is
real; the covariant derivative is $d_A=d+iA$ and the curvature is
$F=dA$ or in components $F_{ij}=\partial_iA_j-\partial_jA_i$.}
\eqn\noneq{\eqalign{F^+_{ij}&=-{i\over 2}\bar M\Gamma_{ij}M \cr
                    \sum_i\Gamma^iD_iM & = 0.\cr}}
In the second equation, $\sum_i\Gamma^iD_i$ is the Dirac operator
$D$ that maps sections of $S^+\otimes L$ to sections of $S^-\otimes L$.
We will sometimes abbreviate the first as $F^+=(M\bar M)^+$.
Alternatively,
if positive spinor indices are written $A,B,C$, and
negative spinor indices as $A',B',C'$,
\foot{Spinor indices are raised and lowered using the invariant
tensor in $\Lambda^2 S^+$.  In components, if $M^A=(M^1,M^2)$,
then $M_A= (-M_2,M_1)$.  One uses the same operation in interpreting
$\bar M$ as a section of $S^+\otimes L$, so $\bar M^A=(\bar M^2,-\bar M^1)$.
Also $F_{AB}={1\over 4}F_{ij}\Gamma^{ij}_{AB}$.}
the equations can be written
\eqn\indeq{\eqalign{F_{AB}& = {i\over 2}\left(M_A\bar M_B+M_B\bar M_A\right)\cr
                    D_{AA'}M^A & = 0.\cr}}

As a first step in understanding these equations, let us work out
the virtual dimension of the moduli space ${\cal M}$
of solutions of the
equations up to gauge transformation.
The linearization of the monopole equations fits into
an elliptic complex
\eqn\pindeq{0\to \Lambda^0\underarrow{s}\Lambda^1
\oplus (S^+\otimes L)\underarrow{t}\Lambda^{2,+}
\oplus (S^-\otimes L) \to 0.}
Here $t$ is the linearization of the monopole equations, and $s$
is the map from zero forms to deformations in $A,M$ given by the infinitesimal
action of the gauge group.  Since we wish to work with real operators
and determine the real dimension
of ${\cal M}$, we temporarily think of $S^\pm\otimes L$ as
real vector bundles (of rank four).
Then an elliptic operator
\eqn\pxxx{T:\Lambda^1\oplus(S^+\otimes L)\to \Lambda^0\oplus \Lambda^{2,+}
\oplus (S^-\otimes L)}
is defined by  $T=s^*\oplus t$.
The virtual dimension of the moduli space is given by the index of $T$.
By dropping terms in $T$ of order zero,
$T$ can be deformed to the direct sum of the operator $d+d^*$
\foot{What is meant here is of course a projection of the $d+d^*$ operator
to self-dual forms.}
from $\Lambda^1$ to $\Lambda^0\oplus \Lambda^{2,+}$ and the Dirac
operator from $S^+\otimes L$ to $S^-\otimes L$.
The index of $T$ is
the index of $d+d^*$ plus twice what is usually called the index of the Dirac
operator; the factor of two comes from looking at $S^{\pm}\otimes L$
as real bundles of twice the dimension.
Let $\chi$ and $\sigma$ be the Euler
characteristic and signature of $X$.  Then the index of $d+d^*$ is
$-(\chi+\sigma)/2$, while twice the Dirac index is $-\sigma/4+c_1(L)^2$.
The virtual dimension of the moduli space is the sum of these or
\eqn\hurf{W=  -{2\chi+3\sigma\over 4} +c_1(L)^2.}

When this number is negative, there are generically no solutions of
the monopole equations.  When $W=0$, that is, when $x=-c_1(L^2)=-2c_1(L)$ obeys
\eqn\burf{x^2=2\chi+3\sigma,}
then the virtual dimension is zero and the moduli space generically
consists of a finite set of points $P_{i,x}$, $i=1\dots t_x$.
With each such point, one can associate
a sign $\epsilon_{i,x}=\pm 1$ -- the sign of the determinant of $T$ as we
discuss momentarily.
Once this is done, define for each $x$ obeying \burf\ an integer $n_x$ by
\eqn\gurofo{n_x=\sum_i\epsilon_{i,x}.}
We will see later that
$n_x=0$ --  indeed, the moduli space is empty -- for all but finitely many $x$.
Under certain conditions that we will discuss in a moment, the $n_x$
are topological invariants.

Note that $W=0$ if and only if the index of the Dirac operator
is
\eqn\inxxon{\Delta={\chi+\sigma\over 4}.}
In particular, $\Delta$ must be an integer to have non-trivial $n_x$.
Similarly, $\Delta$ must be integral for the Donaldson invariants
to be non-trivial (otherwise $SU(2)$ instanton moduli space is odd
dimensional).

For the sign of the determinant of $T$ to make sense one must trivialize
the determinant line of $T$.  This can be done by deforming $T$ as above
to the direct sum of $d+d^*$ and the Dirac operator.  If the Dirac operator,
which naturally has a non-trivial {\it complex} determinant line, is regarded
as a real operator, then its determinant line is naturally trivial -- as a
complex line has a natural orientation.  The $d+d^*$ operator is
independent of $A$ and $M$ (as the gauge group is abelian), and its
deterinant line is trivialized once and for all by picking an orientation
of $H^1(X,{\bf R})\oplus H^{2,+}(X,{\bf R})$.  Note that this is the
same data needed by Donaldson
\ref\donor{S. Donaldson, ``The Orientation Of Yang-Mills Moduli
Spaces And Four-Manifold Topology,'' J. Diff. Geom. {\bf 26} (1987) 397.}
to orient instanton moduli spaces for $SU(2)$;
this is an aspect of the relation between the two theories.

If one replaces $L$ by $L^{-1}$, $A$ by $-A$, and $M$ by $\bar M$, the
monopole equations are invariant, but the trivialization of the
determinant line is multiplied by $(-1)^\delta$ with $\delta$ the Dirac
index.  Hence the invariants for $L$ and $L^{-1}$ are related by
\eqn\pixxx{n_{-x}=(-1)^\Delta n_x.}

For $W<0$, the moduli space is generically empty.  For $W>0$ one can
try, as in Donaldson theory, to define topological invariants that involve
integration over the moduli space.  Donaldson theory does not detect those
invariants at least in known situations.
We will see in section three that even when $W>0$, the
moduli space is empty for almost all $x$.

\bigskip
\noindent{\it Topological Invariance}

In general, the number of solutions
of a system of equations
weighted by the sign of the determinant of the operator analogous to $T$
 is always a topological invariant if a suitable compactness
holds.
If as in the case at hand one has a gauge invariant system of equations, and
one wishes to count gauge orbits of solutions up to gauge transformations,
then one requires (i) compactness; and (ii) free action
of the gauge group on the space of solutions.

Compactness fails if a field or its derivatives can go to
infinity.
The Weitzenbock formula used in section three to discuss vanishing
theorems indicates that these phenomena
do not occur for the monopole equations.
To explain the contrast with Donaldson theory, note that
for $SU(2)$ instantons
compactness fails precisely
\ref\uhl{K. Uhlenbeck, ``Removable Singularities In Yang-Mills Fields,''
Commun. Math. Phys. {\bf 83} (1982) 11.}
because an instanton can shrink to zero size.  This is
possible because (i) the equations are conformally invariant, (ii) they
have non-trivial solutions on a flat ${\bf R}^4$, and (iii) embedding
such a solution, scaled to very small size,
on any four-manifold gives a highly localized approximate
solution of the instanton equations (which can sometimes
\ref\taubes{C. H. Taubes, ``Self-Dual Yang-Mills Connections Over
Non-Self-Dual 4-Manifolds,'' J. Diff. Geom. {\bf 19} (1982) 517.}
be perturbed to
an exact solution).  The monopole equations by contrast
are scale invariant but
(as follows immediately from the Weitzenbock formula) they have
no non-constant $L^2$ solutions on flat ${\bf R}^4$ (or after dimensional
reduction on flat ${\bf R}^n$ with $1\leq n \leq 3$).
So there is no analog for the monopole equations of the phenomenon
where an instanton shrinks to zero size.

On the other hand, an obstruction does arise, just as in Donaldson
theory (in what follows we imitate some arguments in
\ref\dono{S. Donaldson, ``Irrationality And The $h$-Cobordism
Conjecture,'' J. Diff. Geom. {\bf 26} (1987) 141.})
from the question of whether the gauge group acts freely on the
space of solutions of the monopole equations.  The only way for the gauge
group to fail to act freely is that there might be a solution with $M=0$,
in which case a constant gauge transformation acts trivially.
A solution with $M=0$ necessarily has $F^+=0$, that is, it is an abelian
instanton.

Since $F/2\pi$ represents the first Chern class of the line bundle $L$,
it is integral; in particular if $F^+=0$ then $F/2\pi$ lies in the intersection
of the integral lattice in $H^2(X,{\bf R})$ with the anti-self-dual subspace
$H^{2,-}(X,{\bf R})$.
As long as $b_2^+\geq 1$, so that the self-dual part of $H^2(X,{\bf R})$ is
non-empty, the intersection of the anti-self-dual part and the integral
lattice generically consists only of the zero vector.
  In this case,
for a generic metric on $X$, there are no abelian instantons (except for
$x=0$, which we momentarily exclude) and $n_x $ is well-defined.

To show that the $n_x$ are topological invariants, one must further show
that any two generic metrics on $X$ can be joined by a path along which
there is never an abelian instanton.  As in Donaldson theory, this can
fail if $b_2^+=1$.  In that case, the self-dual part
of $H^2(X,{\bf R})$ is one dimensional, and in a generic
one parameter family of metrics on $X$, one may meet a metric for
which there is an abelian instanton. When this occurs, the $n_x$ can jump.
Let us analyze how this happens, assuming for simplicity that $b_1=0$.
Given $b_1=0$ and
$b_2{}^+=1$, one has $W=0$ precisely if the index of the Dirac
equation is 1.  Therefore, there is generically a single solution $M_0$
of the Dirac equation $DM=0$.

The equation $F^+(A)=0$ cannot be obeyed for a generic metric on $X$,
but we want to look at the behavior near a special metric for which it does
have a solution.
  Consider a one parameter family of metrics parametrized
by a real parameter $\epsilon$, such that  at $\epsilon=0$ the
self-dual subspace in $H^2(X,{\bf R})$ crosses a ``wall''
and a solution $A_0$ of
$F^+(A)=0$ appears.  Hence for $\epsilon=0$, we can solve the monopole
equations with $A=A_0, \,M=0$.  Let us see what happens to this solution
when $\epsilon $ is very small but non-zero.  We set $M=mM_0$, with $m$
a complex number, to obey $DM=0$, and we write $A=A_0+\epsilon \delta A$.
The equation $F^+(A)-(M\bar M)^+=0$ becomes
\eqn\nurk{F^+(A_0)+(d\delta A)^+-|m|^2 (M_0\bar M_0)^+=0.}
As the cokernel of $A\to F^+(A)$
is one dimensional, $\delta A$ can be chosen
to project the left hand side of equation \nurk\ into a one dimensional
subspace.  (As $b_1=0$, this can be done in a unique way up to a gauge
transformation.)
The remaining equation looks near $\epsilon=0$ like
\eqn\modlik{c \epsilon -  \,m\bar m=0}
with $c$ a constant.
The $\epsilon$ term on the left comes from the fact that $F^+(A_0)$ is
proportional to $\epsilon$.

Now we can see what happens for $\epsilon\not= 0$ to the solution that
at $\epsilon=0$ has $A=A_0$, $M=M_0$.
Depending on the sign of $c$,
there is a solution for $m$, uniquely
determined up to gauge transformation, for $\epsilon>0$ and no solution
for $\epsilon<0$, or vice-versa.  Therefore $n_x$ jumps by $\pm 1$, depending
on the sign of $c$,
in passing through $\epsilon=0$.
To compare this precisely to the similar behavior of Donaldson
theory, one would also need to understand the
role of the $u$ plane, discussed in section five.

The trivial abelian instanton with $x=0$ is an exception to the
above discussion,
since it cannot be removed by perturbing the metric.  To define $n_0$,
perturb the equation $F_{AB}={i\over 2}(M_A\bar M_B+M_B\bar M_A)$
to
\eqn\hinnoc{F_{AB}={i\over 2}(M_A\bar M_B+M_B\bar M_A)-p_{AB},}
with $p$ a self-dual
harmonic two-form; with this perturbation, the gauge group acts
freely on the solution space.
Then define $n_0$ as the number of gauge orbits of solutions of the
perturbed equations
weighted by sign in the usual way.  This is invariant under continuous
deformations of $p$ for $p\not=0$;
as long as $b_2^+>1$, so that
the space of possible $p$'s is connected, the integer $n_0$ defined
this way is a topological invariant.

The perturbation just
pointed out will be used later in the case that $p$ is the real part
of a holomorphic two-form to compute the invariants of K\"ahler manifolds
with $b_2^+>1$.  It probably has other applications; for instance, the
case that $p$ is a symplectic form is of interest.

\bigskip
\noindent{\it Relation To Donaldson Theory}

With an appropriate restriction on $b_2^+$, the $n_x$ have
(by an argument sketched in section five) a relation to the Donaldson
invariants that will now be stated.

Let us recall that in $SU(2)$ Donaldson theory, one wishes to compute
the integrals or expectation values of certain cohomology classes
or quantum field operators: for every Riemann surface
(or more generally every
two-dimensional homology cycle) $\Sigma$ in $X$, one has an operator
$I(\Sigma)$ of dimension (or $R$ charge or
ghost number) two; there is one additional
operator ${\cal O}$, of dimension four.
For every value of the instanton number, one computes the expectation value
of a suitable product of these operators by integration over instanton
moduli space using a recipe due to Donaldson, or by evaluating a suitable
quantum field theory correlation function as in \witten.
It is natural to organize this data in the form of a generating function
\eqn\jurn{\left\langle
\exp\left(\sum_a\alpha_aI(\Sigma_a)+\lambda {\cal O}\right)
\right\rangle,}
summed over instanton numbers; here
the $\Sigma_a$ range over a basis of $H_2(X,{\bf R})$ and $\lambda,
\,\alpha_a$ are complex numbers.

Let $v= \sum_a\alpha_a[\Sigma_a]$,
with $[\Sigma_a]$ the cohomology class that is Poincar\'e dual to $\Sigma_a$.
So for instance $v^2=\sum_{a,b}\alpha_a\alpha_b\,\,\Sigma_a\cdot\Sigma_b$
(here $\Sigma_a\cdot \Sigma_b$ is the intersection number of $\Sigma_a$ and
$\Sigma_b$), and for any $x\in H^2(X,{\bf Z})$, $v\cdot x=\sum_a\alpha_a
(\Sigma_a,x)$.  Let as before $\Delta=(\chi+\sigma)/4$.

A four-manifold is said to be of simple type if the generating function
in \jurn\ is annihilated by $\partial^2/\partial\lambda^2-4$; all known
simply-connected four-manifolds with $b_2^+>1$ have this property.
The relation of the simple type condition to physics is discussed in
section five.
I claim that for manifolds of simple type
\eqn\jimmo{\eqalign{\left
\langle\exp\left(\sum_a\alpha_aI(\Sigma_a)+\lambda {\cal O}\right)
\right\rangle = 2^{1+{1\over 4}(7\chi+11\sigma)}&\left(\exp\left(
{v^2\over 2}+2\lambda\right)
\sum_x
n_x e^{v\cdot x}\right.\cr&\left.
 +i^{\Delta} \exp\left(-{v^2\over 2}-2\lambda\right)\sum_xn_x
e^{-iv\cdot x}\right).\cr}}
That the expression is real follows from \pixxx.

As sketched in section five, this formula is a sort of corollary of the
analysis of $N=2$ supersymmetric Yang-Mills theory in \sw.  Here I will
just make a few remarks:

(1) The structure in \jimmo\ agrees with the general form
proved by Kronheimer and Mrowka \km.
The classes $x\in H^2(X,{\bf Z})$ such that $n_x\not= 0$ are the basic
classes in their terminology.  From the properties by which $x$ and $n_x$
were defined, we have that $x$ is congruent to $w_2(X)$ modulo 2 and
that $x^2=2\chi+3\sigma$.  The first assertion is a result of Kronheimer
and Mrowka and the second was conjectured by them.

(2) The prefactor $2^{1+{1\over 4}(7\chi+11\sigma)}$ has the following
origin, as in \newwitten.  One factor of two comes because, even though the
center of $SU(2)$ acts trivially on the $SU(2)$ instanton moduli space,
the Donaldson invariants are usually defined without dividing by two.
The remaining factor of $2^{{1\over 4}(7\chi+11\sigma)}$ is a $c$-number
renormalization factor that arises in comparing the microscopic $SU(2)$
theory to the dual description with monopoles.
(In \nsw\ a more general function of the form
$e^{a(u)\chi+b(u)\sigma}$ that arises on the complex $u$ plane will be
calculated.)  Some coefficients in the formula such as the $7/4$ and $11/4$
were fixed in \newwitten\ to agree with
calculations of special cases of Donaldson invariants.

(3) Most fundamentally, in the above formula, the first term, that is
\eqn\kdn{\exp\left({v^2\over 2}+2\lambda\right)
\sum_xn_x e^{v\cdot x},}
is the contribution from one vacuum at $u=\Lambda^2$, and the second
term,
\eqn\hkn{i^\Delta \exp\left(-{v^2\over 2}-2\lambda\right)\sum_xn_x
e^{-iv\cdot x},}
is the contribution of a second vacuum at $u=-\Lambda^2$.
These terms are analogous to the two terms in equation (2.66) of \newwitten.
The factor of $i^\Delta$ arises, as there, because
of a global anomaly in the discrete symmetry that exchanges the two
vacua. This factor
can be written in the form $e^{a\chi+b\sigma}$ and so means
that the two vacua have different values
of the  renormalization mentioned in the last paragraph.  The
replacement of $e^{v\cdot x}$ in the first vacuum by $e^{-iv\cdot x}$
in the second is likewise determined by the symmetries, as in \newwitten,
and can be seen microscopically.
For a general simple compact gauge group, the analogous sum will have
$h$ terms ($h$ the dual Coxeter number) associated with $h$ vacua.

(4) This formula generalizes as follows for the case that the
gauge group is $SO(3)$ rather than $SU(2)$.  Consider an
$SO(3)$ bundle $E$ with, say, $ w_2(E)=z$.
Define a generating functional of correlation functions
summed  over bundles with
 all values of the first Pontryagin class
but fixed $w_2$.  Pick an integral lift of $w_2(X)$, and, using
the fact that the $x$'s are congruent to $w_2(X)$ mod two, let $x'$ be
such that $2x'=x+ w_2(X)$. Then $w_2(E)\not= 0$ modifies
the derivation of \jimmo\ only  by certain minus signs that
appear in the duality transformation that relates the microscopic
and macroscopic descriptions; the result is
\eqn\himmo{
\eqalign{
\left\langle\exp\left(\sum_a\alpha_aI(\Sigma_a)+\lambda {\cal O}\right)
\right\rangle_z
=& 2^{1+{1\over 4}(7\chi+11\sigma)}\left(\exp({v^2\over 2}+2\lambda)
\sum_x(-1)^{x'\cdot z}
n_x e^{v\cdot x} \right.\cr &\left.
+i^{\Delta-z^2} \exp(-{v^2\over 2}-2\lambda)\sum_x
(-1)^{x'\cdot z
}n_xe^{-iv\cdot x}\right).\cr}}
The replacement of $i^\Delta$ by $i^{\Delta-z^2}$ arises, as in equation
(2.79) of \newwitten\ (where $w_2(E)$ is written as $x$),
because the global anomaly has an extra term that depends on $z$.
(Note that as $z$ is defined modulo two, $z^2$ is well-defined modulo four.)
The factor of $(-1)^{x'\cdot z}$ was obtained in \km\ for manifolds
of simple type and in \newwitten\ for K\"ahler manifolds.
If the integral lift of $w_2(X)$ used in defining $x'$
is shifted by $w_2(X)\to w_2(X)+2y$, then \himmo\
is multiplied by $(-1)^{y\cdot z}$.  The reason for this factor
is that \himmo\ is reproducing the conventional Donaldson invariants,
whose sign depends on the orientation of the instanton moduli spaces.
A natural orientation \donor\ depends on an integral lift of $w_2(X)$
and transforms as \himmo\ does if this lift is changed.

(5)  For K\"ahler manifolds with $b_2^+>1$, the quantities entering in \jimmo\
will be completely computed in section four.
We will find that, letting $\eta$ be a holomorphic two-form, the sum in \jimmo\
can be interpreted as a sum over factorizations $\eta=\alpha\beta$
with $\alpha$ and $\beta$ holomorphic sections of $K^{1/2}\otimes L^{\pm 1}$.
Each such factorization contributes $\pm 1$ to $n_x$ with
$x=-2c_1(L)$ provided $x^2=c_1(K)^2$; the contribution is $+1$ or $-1$
according to a formula computed at the end of section four.

\bigskip
\noindent{\it Imitating Arguments From Donaldson Theory}

Apart from relating Donaldson theory to the monopole equations,
one can simply try to adapt familiar arguments about Donaldson theory
to the monopole equations.  We have already seen some examples.

As another example, consider Donaldson's theorem \doninv\
asserting that the Donaldson invariants vanish for a connected sum $X\# Y$
of four-manifolds $X$ and $Y$ which each have $b_2{}^+>0$.  The theorem
is proved by considering a metric on $X\# Y$
in which $X$ and $Y$ are joined by
a long neck of the form ${\bf S}^3\times I$, with $I$ an interval in ${\bf R}$.
Take the metric on the neck to be the product of the standard metric
on ${\bf S}^3$ and a metric that assigns length $t$ to $I$, and consider
the monopole equations on this space.  For $t\to \infty$, any solution
of the monopole equations will vanish in the neck because of the positive
scalar curvature of ${\bf S}^3$ (this follows from the Weitzenbock
formula of the next section).
This lets one define a $U(1)$ action on the moduli space ${\cal M}$
(analogous to the $SO(3)$ action used by Donaldson)
by gauge transforming the solutions on $Y$ by a constant gauge transformation,
leaving fixed the data on $X$.  A fixed point of this $U(1)$ action
would be a solution for which $M$ vanishes on $X$ or on $Y$.  But
as $X$ and $Y$ both have $b_2{}^+>0$, there is no such solution if
generic metrics are used on the two sides.  A zero dimensional moduli
space with a free $U(1)$ action is empty, so the basic invariants would
be zero for such connected sums.  (A free $U(1)$ action also leads
to vanishing of the higher invariants.)
Since we will see in section four
that the invariants are non-zero for K\"ahler manifolds
(analogous to another basic result of Donaldson), one gets a proof
directly from the monopole equations and independent of the equivalence to
Donaldson theory that algebraic surfaces do not have connected
sum decompositions with $b_2^+>0$ on both sides.

If one considers instead a
situation with $b_2^+$ positive for $X$ but zero for $Y$, there will
be fixed points consisting of solutions with $M=0$ on $Y$, and one will get
a formula expressing invariants of $X\# Y$ in terms
of invariants of $X$ and elementary data concerning $Y$.

\newsec{Vanishing Theorems}

Some of the main properties of the monopole equations
can be
understood by means of vanishing theorems.  The general strategy in
deriving such vanishing theorems is quite standard, but as in section two
of \vw, some unusual cancellations (required by the Lorentz invariance
of the underlying untwisted theory) lead to unusually strong results.

If we set $s=F^+-M\bar M$, $k=DM$,
then a small calculation gives
\eqn\highor{\eqalign{\int_Xd^4x\sqrt g\left({1\over 2}|s|^2+|k|^2\right)
=\int_Xd^4x\sqrt g&\left({1\over 2}|F^+|^2+g^{ij}D_iM^AD_j\bar M_A \right.
\cr & \left.+{1\over 2}|M|^4
+{1\over 4}R|M|^2\right) .\cr}}  Here $g$ is the metric of $X$, $R$ the scalar
curvature, and $d^4x\sqrt g$ the Riemannian measure.
A salient feature here is that a term $F_{AB}M^A\bar M{}^B$, which appears
in either $|s|^2 $ or $|k|^2$, cancels in the sum.
This sharpens the implications of the formula, as we will see.
One can also consider the effect here of the perturbation in \hinnoc;
the sole effect of this is to replace
${1\over 2}|M|^4$ in \highor\ by
\eqn\bihor{\int_Xd^4x\sqrt g\left(
F^+\wedge p+\sum_{A,B}\left|{1\over 2}(M_A\bar M_B+M_B\bar M_A)-p_{AB}\right|^2
\right).}
The second term is non-negative, and the first is simply the intersection
pairing
\eqn\juhor{2\pi c_1(L)\cdot [p].}

An obvious inference from \highor\ is that if $X$ admits a metric
whose scalar curvature is positive,
then for such a metric any solution
of the monopole equations must have $M=0$ and $F^+=0$.  But
if $b_2{}^+>0$, then after a generic small perturbation of the metric
(which will preserve the fact that the scalar curvature is positive),
there are no abelian solutions of $F^+=0$ except flat connections.
Therefore,
for such manifolds and metrics, a solution of the monopole equations
is a flat connection with $M=0$.  These too can be eliminated
using the perturbation in \hinnoc.\foot{
Flat connections can only arise if $c_1(L)$ is torsion; in that case,
$c_1(L)\cdot [p]=0$.  The vanishing argument
therefore goes through, the modification in \highor\ being that which
is indicated in \bihor.}
Hence a four-manifold
for which $b_2^+>0$ and $n_x\not= 0$ for some $x$
does not admit a metric of
positive scalar curvature.

We can extend this to determine the possible four-manifolds $X$ with $b_2^+>0$,
some $n_x\not= 0$, and a metric of {\it non-negative}
scalar curvature.\foot{If $b_2^+=1$, the $n_x$ are not all topological
invariants, and we interpret the hypothesis to mean that with at least
one sign of the perturbation in \hinnoc, the $n_x$ are not all zero.}
If $X$ obeys those conditions, then for any metric of $R\geq 0$,
any basic class $x$ is in $H^{2,-}$ modulo torsion
(so that $L$ admits a connection
with $F^+=0$, enabling \highor\ to vanish);
in particular if $x$ is not torsion then $x^2<0$.
Now consider the effect of the perturbation \hinnoc.  As $x\in H^{2,-}$,
\juhor\ vanishes; hence if $R\geq 0$, $R$ must
be zero, $M$
must be covariantly constant and $(M\bar M)^+=p$ (from \bihor).
For $ M$ covariantly constant,
$(M\bar M)^+=p$ implies
that $p$ is covariantly constant also; but for all $p\in H^{2,+}$
to be covariantly constant implies that $X$ is K\"ahler with $b_2^+=1$
or is hyper-K\"ahler.  Hyper-K\"ahler metrics certainly have $R=0$,
and there are examples of metrics with $R=0$
on K\"ahler manifolds with $b_2^+=1$ \ref\lebrun{C. LeBrun, ``Scalar-Flat
K\"ahler Metrics On Blown-Up Ruled Surfaces,'' J. Reine Angew
Math. {\bf 420} (1991) 161.}.

As an example,
for a K\"ahler manifold with $b_2^+\geq 3$, the canonical divisor
$K$ always arises as a basic class, as we will see in section four, so
except in the hyper-K\"ahler case,
such manifolds do not admit a metric of non-negative
scalar curvature.

Even if the scalar curvature is not positive, we can get an explicit
bound from \highor\ showing that there are only finitely many basic classes.
Since
\eqn\gegor{\int_Xd^4x\sqrt g\left({1\over 2}|M|^4+{1\over 4}R|M|^2\right)
\geq -{1\over 32}\int_Xd^4x\sqrt g R^2,}
it follows from \highor, even if we throw away the  term $|D_iM|^2$,
that
\eqn\egor{\int_Xd^4x\sqrt g |F^+|^2\leq {1\over 16}\int_Xd^4x\sqrt g R^2.}
On the other hand, basic classes correspond to line bundles
$L$ with $c_1(L)^2=(2\chi+3\sigma)/4$, or
\eqn\negor{{1\over (2\pi)^2}\int d^4x\sqrt g\left(|F^+|^2-|F^-|^2\right)
        ={2\chi+3\sigma\over 4}.}
Therefore, for a basic class both $I^+=\int d^4x\sqrt g |F^+|^2$
and $I^-=\int d^4x\sqrt g |F^-|^2$ are bounded.  For a given metric,
there are only finitely
many isomorphism classes of line bundles
admitting connections with given bounds on both $I^+$ and $I^-$, so
the set of basic classes is finite.  This is a result
proved by Kronheimer and Mrowka with their definition of the basic classes.

The basic classes correspond, as indicated in section three,
to line bundles on which
the moduli space of solutions of the monopole equations is of zero virtual
dimension.
We can analyze in a similar way components of the moduli space of positive
dimension.  Line bundles $L$ such that $c_1(L)^2<(2\chi+3\sigma)/4$ are not
of much interest in that connection, since for such line bundles the
moduli space has negative virtual dimension and is generically empty.
But if $c_1(L)^2>(2\chi+3\sigma)/4$, then \negor\ is simply replaced by
the stronger bound
\eqn\unegor{{1\over (2\pi)^2}\int d^4x\sqrt g\left(|F^+|^2-|F^-|^2\right)
      >{2\chi+3\sigma\over 4}.}
The set of isomorphism classes of line bundles admitting a connection
obeying this inequality as well as \egor\ is once again finite.
So we conclude that for any given metric on $X$, the set of isomorphism
classes of line bundles for which
the moduli space is non-empty and of non-negative virtual dimension
is finite; for a generic metric on $X$, there are only finitely many
non-empty components of the moduli space.

For further consequences of \highor, we turn to a basic case in the study of
four-manifolds: the case that $X$ is K\"ahler.

\newsec{Computation On K\"ahler Manifolds}

If $X$ is K\"ahler and spin, then $S^+\otimes L$ has a decomposition
$S^+\otimes L\cong (K^{1/2}\otimes L)\oplus (K^{-1/2}\otimes L)$,
where $K$ is the canonical bundle and $K^{1/2}$ is a square root.
If $X$ is K\"ahler but not spin, then $S^+\otimes L$, defined as before,
has a similar decomposition where now $K^{1/2}$ and $L$ are not defined
separately and $K^{1/2}\otimes L$ is characterized
as a square root of the line bundle $K\otimes L^2$.

We denote the components of $M$ in $K^{1/2}\otimes L$ and
in $K^{-1/2}\otimes L$ as $\alpha$ and $-i\bar \beta$, respectively.
The equation $F^+(A)=M\bar M$ can now be decomposed
\eqn\juffy{\eqalign{F^{2,0} & = \alpha\beta \cr
                  F_\omega^{1,1} & =-{\omega\over 2}
                                   \left(|\alpha|^2-|\beta|^2\right)\cr
                   F^{0,2} & =\bar\alpha\bar\beta.\cr}}
Here $\omega$ is the K\"ahler form and $F_\omega^{1,1}$ is the $(1,1)$
part of $F^+$.
\highor\ can be rewritten
\eqn\nohighor{\eqalign{\int_Xd^4x\sqrt g\left({1\over 2}|s|^2+|k|^2\right)
=\int_Xd^4x\sqrt g & \left({1\over 2}|F^+|^2
+g^{ij}D_i\bar\alpha D_j\alpha+g^{ij}D_i\bar\beta
D_j\beta\right.\cr & \left. +{1\over 2}(|\alpha|^2+|\beta|^2)^2
+{1\over 4}R(|\alpha|^2+|\beta|^2)\right) .\cr}}

The right hand side of \nohighor\ is not manifestly non-negative (unless
$R\geq 0$), but the fact that it is equal to the left hand side shows that
it is non-negative and zero precisely for solutions of the monopole
equations.  Consider the operation
\eqn\pohighor{\eqalign{A & \to A\cr
                     \alpha & \to \alpha \cr
                     \beta & \to -\beta.\cr}}
This is not a symmetry of the monopole equations.  But it is a symmetry
of the right hand side of \nohighor.  Therefore, given a zero of the right
hand side of \nohighor\ -- that is, a solution of the monopole equations --
the operation \pohighor\ gives another zero of the right hand side of
\nohighor\ -- that is, another solution of the monopole equations.
So, though not a symmetry of the monopole equations, the transformation
\pohighor\ maps solutions of those equations to other solutions.

Given that any solution of \juffy\ is mapped to another solution by
\pohighor, it follows that such a solution has
\eqn\tohighor{0=F^{2,0}=F^{0,2}=\alpha\beta=\bar\alpha\bar\beta.}
Vanishing of $F^{0,2}$ means that the connection $A$ defines a holomorphic
structure on $L$.
The basic classes (which are first Chern classes of $L$'s that are such that
\juffy\ has a solution) are therefore of type $(1,1)$ for any K\"ahler
structure
on $X$, a severe constraint.

Vanishing of $\alpha\beta$ means that $\alpha=0$
or $\beta=0$.  If $\alpha=0$, then the Dirac equation for $M$ reduces
to
\eqn\jipp{\bar\partial_A \beta=0,}
where $\bar\partial_A$ is the $\bar\partial $ operator on $L$.  Similarly,
if $\beta=0$, then the Dirac equation gives
\eqn\ipp{\bar\partial_A\alpha= 0.}

Knowing that either $\alpha$ or $\beta$ is zero, we can deduce which it is.
Integrating the $(1,1)$ part of \juffy\ gives
\eqn\jippo{{1\over 2\pi}\int_X\omega\wedge F=-{1\over 4\pi}\int_X\omega\wedge
\omega\left(|\alpha|^2-|\beta|^2\right).}
The left hand side of \jippo\ is a topological invariant which can be
interpreted as
\eqn\ippo{J= [\omega]\cdot c_1(L).}
The condition that there are no non-trivial abelian instantons is
that $J$ is non-zero; we only wish to consider metrics for which this
is so.  If $J<0$, we must have $\alpha\not= 0$, $\beta=0$, and if
$J>0$, we must have $\alpha=0$, $\beta\not= 0$.

The equation that we have not considered so far is the $(1,1)$ part of \juffy.
This equation can be interpreted
as follows.  Suppose for example that we are in the situation with $\beta=0$.
The space of connections $A$
and sections $\alpha$ of $K^{1/2}\otimes L$
can be interpreted as a symplectic manifold,
the symplectic structure being defined by
\eqn\defby{\eqalign{\langle\delta_1A,\delta_2A\rangle & =\int_X\omega
          \wedge \delta_1A\wedge\delta_2 A\cr
              \langle \delta_1\alpha,\delta_2\alpha
\rangle & =-i\int_X\omega\wedge\omega
\left(\delta_1\overline \alpha\delta_2\alpha-\delta_2\bar \alpha
\delta_1\alpha\right).\cr}}
On this symplectic manifold acts the group of $U(1)$ gauge transformations.
The moment map $\mu$ for this action is the quantity that appears in the
$(1,1)$ equation that we have not yet exploited, that is
\eqn\hefby{ \mu\omega= F_\omega^{1,1}+\omega|\alpha|^2.}
By analogy with many similar problems, setting
to zero the moment map and dividing by the group of $U(1)$ gauge
transformations
should be
equivalent to dividing by the complexification of the group of gauge
transformations.\foot{In such comparisons of symplectic and complex quotients,
one usually needs a stability condition on the complex side.
In the present case, this is the condition discussed in
connection with \ippo.}  In the present case, the complexification of the
group of gauge transformations acts by $\alpha\to t\alpha$,
$\bar\partial_A\to t\bar\partial_At^{-1}$, where $t$ is a map from
$X$ to ${\bf C}^*$.

Conjugation by $t$ has the effect of identifying any two $A$'s that
define the same complex structure on $L$.  This can be done almost
uniquely: the ambiguity is that conjugation by a constant $t$ does
not change $A$.  Of course, a gauge transformation by
a constant $t$ rescales
$\alpha$ by a constant.  The result therefore, for $J<0$, is that the moduli
space of solutions of the monopole equations is the moduli space of
pairs consisting of a complex structure on $L$ and a non-zero
holomorphic section, defined
up to scaling, of $K^{1/2}\otimes L$.  For $J>0$, it is instead
$\beta$ that is non-zero, and $K^{1/2}\otimes L$ is replaced by
$K^{1/2}\otimes L^{-1}$.

This result can be stated particularly nicely if $X$ has $b_1=0$.
Then the complex structure on $L$, assuming that it exists, is unique.
The moduli space of solutions of the monopole equations is
therefore simply a complex projective space, ${\bf P}H^0(X,K^{1/2}\otimes L)$
or ${\bf P}H^0(X,K^{1/2}\otimes L^{-1})$, depending on the sign of $J$.

\bigskip
\noindent{\it Identifying The Basic Classes}

We would now like to identify the basic classes.
The above description of the moduli space gives considerable information:
basic classes are of the form $x=-2c_1(L)$, where $L$ is such that
$J<0$ and $H^0(X,K^{1/2}\otimes L)$ is non-empty, or $J>0$
and $H^0(X,K^{1/2}\otimes L^{-1})$ is non-empty.  This, however,
is not a sharp result.

That is closely related to the fact that the moduli spaces ${\bf P}H^0(X,
K^{1/2}\otimes L^{\pm 1})$ found above very frequently have a dimension
bigger than the ``generic'' dimension of the moduli space as predicted
by the index theorem.  In fact, K\"ahler metrics
are far from being generic.  In case the expected dimension
is zero, one would have always $n_x>0$ if the moduli spaces behaved
``generically'' (given the complex orientation, an isolated point on the
moduli space would always contribute $+1$ to $n_x$; this is a special
case of a discussion below).  Since the $n_x$
are frequently negative (as in the examples of Kronheimer
and Mrowka or equation (2.66) of \newwitten), moduli spaces of
non-generic dimension must appear.

When the moduli space has greater than the generically expected dimension,
one can proceed by integrating over
the bosonic and fermionic collective
coordinates in the path integral.  This gives a result that can be
described topologically: letting $T$ be the operator that arises in linearizing
the monopole equations, the cokernel of $T$ is a vector bundle $V$
(the ``bundle of antighost zero modes'') over the moduli space ${\cal M}$;
its Euler class integrated over ${\cal M}$ is the desired $n_x$.

Alternatively, one can perturb the equations to more generic ones.
We use the same perturbation as before.
For a K\"ahler manifold $X$, the condition $b_2^+>1$ is equivalent
to $H^{2,0}(X)\not= 0$, so we can pick a non-zero holomorphic two-form
$\eta$.\foot{In \newwitten, where essentially the same perturbation
was made, the two-form was called $\omega$, but
here we reserve that name for the K\"ahler form.}
We perturb the monopole equations \juffy\
to
\eqn\ojuffy{\eqalign{F^{2,0} & = \alpha\beta -\eta\cr
            F_\omega^{1,1} & = -\omega\left(|\alpha|^2-|\beta|^2\right)\cr
                   F^{0,2} & =\bar\alpha\bar\beta-\bar\eta,\cr}}
leaving unchanged the Dirac equation for $M$.

It suffices to consider the case that the first Chern class of $L$
is of type $(1,1)$, since the unperturbed moduli space vanishes otherwise.
That being so, we have
\eqn\ijuffy{0=\int_XF^{2,0}\wedge\bar\eta=\int_XF^{0,2}\wedge \eta.}
Using this, one finds that \nohighor\ generalizes to
\eqn\onohighor{\eqalign{\int_Xd^4x\sqrt g\left({1\over 2}|s|^2+|k|^2\right)
=\int_Xd^4x &\left({1\over 2}
|F^+|^2+g^{ij}D_i\bar\alpha D_j\alpha+g^{ij}D_i\bar\beta
D_j\beta\right.\cr & \left.
 +{1\over 2}(|\alpha|^2-|\beta|^2)^2+{2}|\alpha\beta-\eta|^2
+{R\over 4}(|\alpha|^2+|\beta|^2)\right) .\cr}}
We can now make an argument of a sort we have already seen: the transformation
\eqn\hoggy{\eqalign{A & \to A\cr
                    \alpha & \to \alpha \cr
                    \beta  & \to -\beta \cr
                    \eta   & \to -\eta,  \cr}}
though not a symmetry of \ojuffy, is a symmetry of the right hand side of
\onohighor.  As solutions of \ojuffy\ are the same as zeroes of the right
hand side of \onohighor, we deduce that the solutions of \ojuffy\ with
a two-form $\eta$ are transformed by \hoggy\ to the solutions with $-\eta$.
The terms in \ojuffy\ even or odd under the transformation must therefore
separately vanish, so
any solution of \ojuffy\ has
\eqn\goggy{0= F^{0,2}=F^{2,0}=\alpha\beta-\eta.}
The condition $F^{0,2}=0$ means that the connection still defines
a holomorphic structure on $L$.

The condition
\eqn\jipoggy{ \alpha\beta =\eta}
gives our final criterion for determining the basic classes: they are
of the form
$x=-2c_1(L)$  where, for any choice of $\eta\in H^0(X,K)$, one has
a factorization $\eta=\alpha\beta$ with
holomorphic sections $\alpha$ and $\beta$ of $K^{1/2}\otimes L^{\pm 1}$,
and $x^2=c_1(K)^2$.

To make this completely explicit, suppose
the divisor of $\eta$ is a union of irreducible
components $C_i$ of multiplicity $r_i$.
Thus the canonical divisor is
\eqn\rufu{c_1(K)=\sum_ir_i[C_i],}
where $[C_i]$ denotes the cohomology class that is
Poincar\'e dual to the curve $C_i$.
The existence of the factorization $\eta=\alpha\beta$
means that the divisor of $K^{1/2}\otimes L$ is
\eqn\jufu{c_1(K^{1/2}\otimes L)=\sum_is_i[C_i],}
where $s_i$ are integers with $0\leq s_i\leq r_i$.
The first Chern class of $L$ is therefore
\eqn\tufu{c_1(L)=\sum_i(s_i-{1\over 2}r_i)[C_i].}
And the basic classes are of the form $x=-2c_1(L)$ or
\eqn\pufu{x=-\sum_i(2s_i-r_i)[C_i].}

An $x$ of this form is is of type $(1,1)$ and congruent to $c_1(K)$
modulo two, but may not obey $x^2=c_1(K)^2$.
It is actually possible to prove using the Hodge index theorem
that for $x$ as above, $x^2\leq c_1(K)^2$.\foot{Such an argument
was pointed out by D. Morrison.}  This is clear from the monopole
equations: perturbed to $\eta\not=0$, these equations have
 at most isolated solutions
(from the isolated factorization $\eta=\alpha\beta$) and not a moduli
space of solutions of positive dimension.  So for K\"ahler manifolds,
the non-empty perturbed moduli spaces are at most of dimension zero; invariants
associated with monopole moduli spaces of higher dimension vanish.

Our final conclusion about the basic classes, then, is that they
are classes of the form \pufu\ such that $x^2=c_1(K)^2$.
Each factorization $\eta=\alpha\beta$ contributes
$\pm 1$ to $n_x$ with the corresponding $x$.
Since several factorizations might give the same $x$, cancellations
may be possible, making it possible to write the invariant in
the Kronheimer-Mrowka form, with a shorter list of basic classes.
Such cancellations can be effectively found since the signs of the various
contributions are computed below.  In any event the classes $x=\pm K$
arise only from $s_i=0$ or $s_i=r_i$, respectively, and so always
arise as basic classes with $n_x=\pm 1$.
\foot{G. Tian and S.-T. Yau, P. Kronheimer and T. Mrowka, D. Morrison,
and R. Friedman and J. Morgan pointed out that it actually follows
from these conditions (or related arguments)
that if $X$ is a minimal surface of general
type, then the only basic classes are $\pm K$ (so that $K$ is a differentiable
invariant up to sign).  Indeed, according to Lemma 4 in \ref\kodaira{
K. Kodaira, ``Pluricanonical Systems On Algebraic Surfaces Of General
Type,'' J. Math. Soc. Japan {\bf 20} (1968) 170.}, on such a surface, if
$K={\cal O}(C_1)\otimes {\cal O}(C_2)$
with non-zero effective divisors $C_1,C_2$, then $C_1\cdot C_2>0$.
This means that a factorization $\eta=\alpha\beta$ with $\alpha,\beta$
sections of $K^{1/2}\otimes L^{\pm 1}$ and $x^2=c_1(K)^2$
implies that $K^{1/2}\otimes L^{\pm 1}$ is trivial with one choice of sign,
and hence that $x=\pm c_1(K)$.}

\bigskip
\noindent{\it Comparison To Previous Results}

Let us compare these statements to previous results.  The main case considered
in \newwitten\ was that in which the $C_i$ were disjoint
with multiplicities $r_i=1$.  The allowed values of the $s_i$ are
then $0$ and 1, so the basic classes are
\eqn\ppufu{x_{\vec \rho}=\sum_i\rho_i[C_i],}
with each $\rho_i=\pm 1$, as claimed in \newwitten.
Notice that all of these classes have $x_{\vec\rho}^2=c_1(K)^2$.

The most important case in which the $r_i$ are not all one is the case
of an elliptic surface with multiple fibers.  A fiber of multiplicity $n$
appears in the canonical divisor with weight $r=n-1$.  For elliptic
surfaces, one has $C_i\cdot C_j=0$ for all $i,j$,
so the classes in \pufu\ actually do all have
$x^2=c_1(K)^2=0$.  The formulas of
Kronheimer and Mrowka for the Donaldson invariants of these surfaces show
that the basic classes, in their sense,
are indeed the classes given in \pufu.

\bigskip
\noindent{\it Determination Of The Sign}

To complete this story, we must compute, for each factorization,
the sign of $\det T$.
Let us first explain in an abstract setting the strategy that will be
used.
Suppose that $E$ and $F$ are real vector spaces of equal even dimension
with given complex structures, and $T:E\to F$ is an invertible linear map that
commutes with the complex structure.  Then $\det T$ is naturally
defined as an element of $\det F\otimes \det E^{-1}$.
If $\det E$ and $\det F$ are trivialized using the complex orientations of $E$
and $F$, then $\det T>0$ roughly because the complex structure gives a pairing
of eigenvalues.  If $T$ {\it reverses} the complex structures then
the sign of $\det T$ is $(-1)^w$ with $w=\dim_{\bf C}E$.  For instance,
by reversing the complex structure of $E$ one could reduce to the case
in which $T$ preserves the complex structures, but reversing the complex
structure of $E$ multiplies its orientation by $(-1)^w$.

One can combine the two cases as follows.  Suppose that $T$ preserves
the complex structures but is not invertible.  Let $T':E\to F$
be a map that reverses the complex structures and maps ${\rm ker}\,T$
invertibly to $F/T(E)$.  Then for small real $\epsilon$ (of any sign)
the sign of $\det(T\oplus \epsilon T')$ is $(-1)^w$ where now
$w={\rm dim}_{\bf C}{\rm ker}\,T$.  The same formula holds if
$U$ and $V$ are vector bundles,
$E=\Gamma(U)$, $F=\Gamma(V)$, $T:E\to F$
is an elliptic operator with zero index, $T'$ is a sufficiently mild
perturbation, and $\det (T+\epsilon T')$ is
understood as the Ray-Singer-Quillen
determinant.

Our problem is of this form with $T$ understood as the linearization
of the monopole equations at $\eta=0$ and $T'$ as the correction
proportional to $\eta$ (which enters the linearization because of the shift
it induces in $\alpha$ or $\beta$).
As in \pxxx, one has $U=\Lambda^1\oplus (S^+\otimes L)$,
with $S^+\otimes L$ now regarded as a real vector bundle of rank four.
If $J<0$ (so $\beta=0$ for $\eta=0$), then
give $U$ a complex structure that acts naturally on
$S^+\otimes L$ and multiplies $\Lambda^{0,1}$ and $\Lambda^{1,0}$ by
$i$ and $-i$, respectively.  Likewise
give $V=\Lambda^0\oplus\Lambda^{2,+}\oplus (S^-\otimes L)$
a complex structure that acts naturally on $S^-\otimes L$; multiplies
$\Lambda^{0,2}$ and $\Lambda^{2,0}$ by $i$ and $-i$; and exchanges the
$(1,1)$ part of $\Lambda^{2,+}$ with $\Lambda^0$.
Then $T$ preserves the
complex structures on these bundles and $T'$ reverses them.

The sign of the contribution to $n_x$ from a factorization $\eta=\alpha\beta$
is therefore $(-1)^w$ with $w={\rm dim}_{\bf C}{\rm \ker}\,T$.
The kernel of $T$ can be described as follows.  There is an exact
sequence
\eqn\immo{0\to {\cal O}\underarrow{\alpha}K^{1/2}\otimes L\to R\to 0,}
with some sheaf $R$.  The kernel of $T$ has the same dimension as
$H^0(X,R)$, as explained below.  So the sign of the contribution to $n_x$
is
\eqn\uddu{(-1)^{{\rm dim}\,H^0(X,R)}.}
If instead $J>0$, so the unperturbed solution has $\alpha=0$, $\beta\not=0$,
then first of all we reverse the complex structures on $S^\pm\otimes L$;
this multiplies the determinant by $(-1)^\Delta$ where $\Delta=-\sigma/8
+c_1(L)^2/2=(\chi+\sigma)/4$ is the Dirac index.  The rest of
the discussion goes through with
\immo\ replaced by
\eqn\dimmo{0\to {\cal O}\underarrow{\beta}K^{1/2}\otimes L^{-1}
\to \tilde R\to 0,}
so the sign is
\eqn\duddu{(-1)^{\Delta+{\rm dim}\,H^0(X,\tilde R)}.}
(It can be verified using the classification of surfaces that \uddu\
and \duddu\ are equal.)
With these signs, \jimmo\ becomes completely explicit: the sum in \jimmo\ is
a sum over factorizations $\eta=\alpha\beta$; each such factorization
determines a class $x$ and contributes to $n_x$ an amount $\pm 1$
as just determined.

Before justifying the claim about $\ker T$, let us
check that the sign just determined agrees with what has been computed
by other methods.  Suppose as in \newwitten\ that the divisor of $\eta$
is a union of disjoint smooth curves $C_i$. Then $R$ is a sum of
sheaves $R_i$ supported on $C_i$; $R_i$ is trivial if $s_i$
(defined in \jufu) is 0 and is isomorphic to a  spin bundle of $C_i$
(determined by $\eta $ and independent of the factorization $\eta=\alpha\beta$)
if $s_i=1$.
Let $t_i=1$ if this spin bundle is even, that is, if ${\rm dim}\,H^0(C_i,R_i)$
is even, and $-1$ if it is odd.  Then \uddu\ becomes
\eqn\nurfo{(-1)^{{\rm dim}\,H^0(X,R)}=\prod_{i|s_i=1} t_i.}
This is the result claimed in equation (2.66)
of \newwitten.  One can similarly check that \jimmo\ when evaluated
with the signs given above agrees with the formulas of Kronheimer
and Mrowka for Donaldson invariants of elliptic surfaces with multiple
fibers.

It remains to justify the claimed structure of $\ker\, T$.  Suppose, for
instance, that
we are linearizing around a solution with $\beta=0$, $\alpha\not= 0$.
Let $\delta A$, $\delta \alpha$, and $\delta \beta$ denote
first order variations of $A,\alpha,$ and $\beta$.  The argument
that proves the vanishing theorem shows that for $\delta A,\delta\alpha,
\delta\beta$ to be annihilated by $T$, one must
have $\alpha\delta\beta=0$ and hence $\delta\beta=0$.  The remaining
equations can be written
\eqn\remeq{\eqalign{\bar\partial \,\,\delta A^{0,1} & = 0 \cr
                     i\delta A^{0,1}\alpha +\bar\partial_A\delta\alpha & = 0.
\cr}}
One must divide the space of solutions of \remeq\ by solutions that arise
from complex gauge transformations of $A,\alpha$.
If $\delta A^{0,1}=0$, then the second equation says that $\delta\alpha
\in H^0(X, K^{1/2}\otimes L)$; however, upon dividing by complex
gauge transformations (which include rescalings of $\alpha$ by a constant)
we should regard $\delta\alpha$ as an element of $H^0(X,K^{1/2}\otimes L)/
{\bf C}\alpha$.   The first equation says that $\delta A^{0,1}$ defines
an element of $H^1(X,{\cal O})$, and the second equation says that
multiplication by $\alpha$ maps this element to zero in $H^1(X,K^{1/2}\otimes
L)$.  So if ${\rm ker}\,\alpha$ is the kernel of
$H^1(X,{\cal O})\underarrow{\alpha}H^1(X,K^{1/2}\otimes L)$, then there
is an exact sequence
\eqn\imoc{0\to H^0(X,K^{1/2}\otimes L)/{\bf C}\alpha
\to {\rm ker}\,T\to {\rm ker}\,\alpha\to
0.}
This can be compared to the exact sequence
\eqn\nimoc{0\to H^0(X,K^{1/2}\otimes L)/{\bf C}\alpha
\to H^0(X,R)\to {\rm ker}\,\alpha\to
0}
that comes from \immo.  Comparison of these sequences shows
that ${\rm ker}\,T$ and $H^0(X,R)$ have the same dimension,
as asserted above; one should be able to identify these spaces canonically.

\newsec{A Short Sketch Of The Physics}

To sketch the relation of these ideas to quantum field theory,
let us first recall the analysis of $N=2$ supersymmetric Yang-Mills
theory in \sw.  To begin with we work on flat ${\bf R}^4$.
It has long been known that this theory has a family of quantum vacuum
states parametrized by a complex variable $u$, which corresponds
to the four dimensional class in Donaldson theory.  For $u\to\infty$,
the gauge group is spontaneously broken down to the maximal torus,
the effective coupling is small, and everything can be computed
using asymptotic freedom.  For small $u$, the effective coupling is strong.
Classically, at $u=0$, the full $SU(2)$ gauge symmetry is restored.
But the classical approximation is not valid near $u=0$.

Quantum mechanically, as explained in \sw, the $u$ plane turns out to
parametrize a family of elliptic curves,
\foot{If $SU(2)$ is replaced by
a Lie group of rank $r$, elliptic curves are replaced
by abelian varieties of rank $r$; the analog of
the simple type condition is that the commutative
algebra of operators obtained by evaluating the Chern classes of the universal
bundle at a point in a four-manifold has a spectrum consisting of $h$
points ($h$, which is $N$ for $SU(N)$, is the dual Coxeter number of the
Lie group) where these varieties degenerate maximally.}
in fact, the modular curve
of the group $\Gamma(2)$.  The family can be described by the equation
\eqn\urmo{y^2=(x^2-\Lambda^4)(x-u),}
where $\Lambda$ is the analog of a parameter that often goes by the same
name in the theory of strong interactions.  (The fact that $\Lambda\not= 0$
means that the quantum theory does not have the conformal invariance
of the classical theory.)
The curve \urmo\ is smooth for generic $u$, but degenerates to
a rational curve for $u=\Lambda^2,-\Lambda^2$, or $\infty$.  Near each
degeneration, the theory becomes weakly coupled, and everything is calculable,
if the right variables are used.  At $u=\infty$, the weak coupling is
(by asymptotic freedom) in terms of the original field
variables.  Near $u=\pm \Lambda^2$
a magnetic monopole (or a dyon, that is a particle carrying both
electric and magnetic charge) becomes massless; the light degrees
of freedom are the monopole or dyon and a dual photon or $U(1)$ gauge
boson.  In terms of the dyon and dual photon, the theory is weakly
coupled and controllable near $u=\pm \Lambda^2$.

Notice that quantum mechanically on flat ${\bf R}^4$,
the full $SU(2)$ gauge symmetry is never  restored.  The only really
exceptional behavior that
occurs anywhere is that magnetically charged particles
become massless.

Now, for any $N=2$ supersymmetric field theory, a standard twisting
procedure \witten\ gives a topological field theory.  In many cases,
these topological field theories are related to the counting of
solutions of appropriate equations.   For instance, the procedure,
applied to the underlying $SU(2)$ gauge theory, gives Donaldson theory
(that is, the problem of counting $SU(2)$ instantons); applied to
the quantum theory near $u=\pm \Lambda^2$, it gives the problem of
counting the solutions of the monopole
equations; applied at a generic point on the $u$ plane, it gives, roughly,
the problem of counting {\it abelian} instantons.

Now let us apply this experience, to work
on a general oriented four-manifold $X$.
The structure of the argument is analogous to the heat kernel proof
of the index theorem, in which one considers the trace of the heat kernel
$\Tr (-1)^Fe^{-tH}$.  This is independent of $t$ but can be evaluated
in different ways for $t\to 0$ or for $t\to \infty$;
for small $t$, one sees local geometry and gets
a cohomological formula, while for large $t$, one gets a description
in terms of the physical ground states (harmonic spinors).

In the four-manifold problem, letting $g$ be any Riemannian
metric on $X$, we consider the one parameter family of metrics $g_t=tg$,
with $t>0$.   Correlation functions of the twisted topological field
theory are metric independent and so independent of $t$.
For $t\to 0$, using asymptotic freedom,
the classical description becomes valid,
and one recovers Donaldson's definition of four-manifold invariants
from the $N=2 $ theory.  In particular, for four-manifolds on which
there are no abelian instantons, the main contribution comes from
$u=0$ where for small $t$ one computes in the familiar fashion
with the full $SU(2)$ gauge theory.

What happens for large $t$?  Once the scale of the four-manifold
is much greater than $1/\Lambda$, the good description is in terms of
the degrees of freedom of the vacuum states on ${\bf R}^4$.  At first
sight, it might appear that the answer will come by integration over
the $u$ plane.  That is apparently so for some classes of problems.

However, for four-manifolds with $b_2^+>1$, one can show
that the contribution of any region of the $u$ plane bounded away
from $u=\pm \Lambda^2$ vanishes as a power of $t$ for $t\to \infty$.
This is roughly because in the abelian theory that prevails away
from $u=\pm \Lambda^2$, there are too many fermion zero modes
and no sufficiently efficient way to lift
them.  (It is not clear if the gap in the
argument for non-K\"ahler manifolds with $b_2^+=3$ is significant,
or could be removed with a more precise treatment.)

Under the above condition on $b_2^+$, a contribution that survives
for $t\to\infty$ can therefore come only from a neighborhood of
$u=\pm \Lambda^2$ that shrinks to zero as $t$ grows.  The contribution
from this region does survive for $t\to \infty$; it can be computed
using the monopole equations since those are the relevant equations
in the topologically twisted theory near $u=\pm \Lambda^2$.
In computing a correlation function of operators of the twisted theory
near $u=\pm \Lambda^2$, one can expand all operators of the microscopic
theory in terms of operators of successively higher dimension in the
macroscopic, monopole theory.

For $u$, the most relevant term (that is, the term of lowest dimension)
is the $c$-number $u=\Lambda^2$ or $u=-\Lambda^2$.
The simple type condition -- which asserts that $u$ is semi-simple
with a spectrum consisting of two points -- arises when one may
replace $u$ by this $c$-number.  For the operator
related to the  two-dimensional classes of Donaldson theory, the most
relevant term is again a $c$-number,
measuring the first Chern class of the dual line bundle $L$ of the monopole
problem. Keeping only these terms, since the operators are replaced
by    $c$-numbers, correlation
functions can be computed by simply
counting solutions weighted by the sign of the fermion determinant; only zero
dimensional moduli spaces contribute.  Upon fixing the normalizations
by comparing to known special cases, one arrives at \jimmo.

This in fact appears to be justified since as usual in such
problems operators of higher
dimension give contributions that vanish as negative powers of $t$.
This would give a quantum field theory proof that
all oriented four-manifolds with $b_2^+>3$ are of simple type.
If, however, higher terms in the expansion of the operators survive
on some four-manifolds with $b_2^+>3$, the consequences would be as follows.
Then the higher monopole invariants of $W\not= 0$ can be detected in
Donaldson theory, and \jimmo\ will be replaced by a more general
formula involving the expansion near $u=\pm \Lambda^2$ of some of the
functions computed in \sw.  The number $s$ of higher terms that one would have
to keep in the expansion would be one half
the maximum value of $W$ that contributes.  $u$ will still have a spectrum
consisting of two points, but instead of $u^2-\Lambda^4=0$, one would
get $(u^2-\Lambda^4)^{s+1}=0$.  Such a situation has in fact been
analyzed by Kronheimer and Mrowka.

\listrefs
\end